\documentclass[11pt]{article} 
\usepackage{amsmath}
\usepackage{amsfonts,amssymb}

\begin{document}

\newcommand{\nl}{\nonumber\\}
\newcommand{\nnl}{\nl[6mm]}
\newcommand{\nle}{\nl[-2.0mm]\\[-2.0mm]}
\newcommand{\nlb}[1]{\nl[-2.0mm]\label{#1}\\[-2.0mm]}
\newcommand{\ab}{\allowbreak}

\renewcommand{\leq}{\leqslant}
\renewcommand{\geq}{\geqslant}

\renewcommand{\theequation}{\thesection.\arabic{equation}}
\let\ssection=\section
\renewcommand{\section}{\setcounter{equation}{0}\ssection}

\newcommand{\be}{\bes}
\newcommand{\ee}{\ees}
\newcommand{\bes}{\begin{eqnarray}}
\newcommand{\ees}{\end{eqnarray}}
\newcommand{\eens}{\nonumber\end{eqnarray}}

\renewcommand{\/}{\over}
\renewcommand{\d}{\partial}

\newcommand{\mm}{{\mathbf m}}
\newcommand{\nn}{{\mathbf n}}
\newcommand{\phim}{\phi_{,\mm}}
\newcommand{\phin}{\phi_{,\nn}}
\newcommand{\pim}{{\pi}^{,\mm}}
\newcommand{\pin}{{\pi}^{,\nn}}

\newcommand{\si}{\sigma}
\newcommand{\eps}{\epsilon}
\newcommand{\dlt}{\delta}
\newcommand{\om}{\omega}
\newcommand{\al}{\alpha}
\newcommand{\vth}{\vartheta}
\renewcommand{\th}{\theta}
\newcommand{\rep}{\varrho}

\newcommand{\xmu}{\xi^\mu}
\newcommand{\dmu}{\d_\mu}
\newcommand{\dnu}{\d_\nu}

\newcommand{\ssg}{sl(3)\!\oplus\! sl(2)\!\oplus\! gl(1)}

\newcommand{\no}[1]{{\,:\kern-0.7mm #1\kern-1.2mm:\,}}

\newcommand{\vect}{{\mathfrak{vect}}}
\newcommand{\svect}{{\mathfrak{svect}}}
\newcommand{\map}{{\mathfrak{map}}}
\newcommand{\mb}{{\mathfrak{mb}}}
\newcommand{\ksle}{{\mathfrak{ksle}}}
\newcommand{\vle}{{\mathfrak{vle}}}
\newcommand{\kas}{{\mathfrak{kas}}}
\newcommand{\vas}{{\mathfrak{vas}}}
\newcommand{\kk}{{\mathfrak{k}}}
\newcommand{\fle}{{\mathfrak {le}}}
\newcommand{\sle}{{\mathfrak {sle}}}
\newcommand{\ko}{{\mathfrak {m}}}
\newcommand{\sko}{{\mathfrak {sm}}}

\renewcommand{\L}{{\cal L}}
\newcommand{\Lxi}{\L_\xi}

\newcommand{\im}{{\rm im}}
\renewcommand{\div}{{\rm div}}
\newcommand{\afn}{{\rm afn\,}}
\newcommand{\til}{{\tilde{\ }}}

\newcommand{\fa}{\phi_\alpha}
\newcommand{\Ea}{\EE^\alpha}
\newcommand{\fs}{\phi^*}
\newcommand{\fsa}{\phi^{*\alpha}}
\newcommand{\psa}{\pi^*_\alpha}

\newcommand{\qmu}{q^\mu}
\newcommand{\pmu}{p_\mu}
\newcommand{\pnu}{p_\nu}

\newcommand{\ud}{u^i\d_i}
\newcommand{\thd}{\th_{ia}d^{ia}}
\newcommand{\veth}{\vth^a\eth_a}

\newcommand{\larroww}[1]{{\ \stackrel{#1}{\longleftarrow}\ }}
\newcommand{\brep}[1]{{\bf{\underline{#1}}}}

\newcommand{\tr}{{\rm tr}\kern0.7mm}
\newcommand{\oj}{{\mathfrak g}}
\newcommand{\hh}{{\mathfrak h}}
\newcommand{\U}{{\cal U}}
\newcommand{\QQ}{{\cal Q}}
\newcommand{\EE}{{\cal E}}
\newcommand{\II}{{\cal I}}
                                            
\newcommand{\TT}{{\mathbb T}}
\newcommand{\RR}{{\mathbb R}}
\newcommand{\CC}{{\mathbb C}}
\newcommand{\ZZ}{{\mathbb Z}}
\newcommand{\NN}{{\mathbb N}}

\title{{Symmetries of Everything}}

\author{T. A. Larsson \\
email: thomas.larsson@hdd.se}

\maketitle
\begin{abstract}
I argue that string theory can not be a serious candidate for the 
Theory of Everything, not because it lacks experimental support, but
because of its algebraic shallowness. I describe two classes of algebraic 
structures which are deeper and more general than anything seen in 
string theory: 
\begin{enumerate}
\item
The multi-dimensional Virasoro algebras, i.e. the 
abelian but non-central extension of the algebra of vector fields in 
$N$ dimensions by its module of closed dual one-forms. 
\item
The exceptional simple Lie superalgebra $\mb(3|8)$, which is the 
deepest possible symmetry (depth $3$ in its consistent Weisfeiler 
grading). The grade zero subalgebra, which largely governs the 
representation theory, is the standard model algebra $\ssg$.
Some general features can be extracted from an $\mb(3|8)$ gauge theory 
even before its detailed construction: several generations of fermions,
absense of proton decay, no additional gauge bosons, manifest CP
violation, and particle/anti-particle asymmetry.
\end{enumerate}
I discuss classifications supporting the claim that every 
conceivable symmetry is known.
\end{abstract}

\section{What's wrong with string theory?}

Recently\footnote{The bulk text was written in spring 2001 and is 
identical to version 2, except for the corrected definition of $\mb(3|8)$,
which has been available in \cite{Lar01b}. The footnotes contain some 
observations added after the appearance of D. Friedan: {\it A tentative
theory of large-distance physics}, {\tt hep-th/0204131} (2002). }
it has become clear that not all mathematical physicists
are entirely convinced that string theory is the ultimate Theory of 
Everything (ToE) \cite{Sch99,Woit01}\footnote{Note added in version 3:
A popular version of Woit's article appeared in American Scientist
{\bf 90} 110--112 (March-April 2000).}. According to my own prejudices, 
there are only two valid reasons to pursue a branch of theoretical 
physics: 
experimental support and intrinsic mathematical beauty. String theory has 
always disagreed with experiments, to the extent that it makes any 
falsifyable predictions whatsoever (extra dimensions, supersymmetric
partners, $496$ gauge bosons, ...). Of course, one may always 
argue that these features will show up at higher energies, not accessible 
to experimental falsification within the forseeable future.
However, my objection to string theory is not the fact that it is
experimentally wrong (or void, according to taste), but rather that it 
is not based on the deepest and most general algebraic structures 
conceivable\footnote{It is difficult to argue about mathematical beauty,
since beauty lies in the eyes of the beholder. Generality is
less ambiguous, and depth even has a technical definition, although I 
admit that using this term to mean profound is a rhetoric trick.}.
The algebraic structures of string theory (finite-dimensional semisimple 
Lie algebras such as $so(32)$ and $E_8\times E_8$,
and central extensions of infinite-dimensional Lie algebras
such as the Virasoro and superconformal algebras) were probably the 
deepest structures known in 1984, at the time of the first string 
revolution. Actually, several of 
these algebras were discovered by string theorists in an earlier era. 
However, the algebra community has not been lazy after 1984. During the 
1990s, there has been progress along at least two lines, which make the
symmetries of string theory appear shallow in comparison. 

1. String theory relies on the distinguished status of central extensions 
of infinite-dimensional Lie (super)algebras of linear 
growth. While it is true that only a few algebras admit 
central extensions, the algebra $\vect(N)$ of vector fields acting on 
$N$-dimensional spacetime admits abelian but non-central extensions, 
which naturally generalize the Virasoro algebra to $N>1$ 
dimensions\footnote{Privately, I refer to the abelian but non-central 
Virasoro-like cocycles as {\em viral cocycles}.}. 
Geometrically, the multi-dimensional Virasoro 
algebra is an extension by the module of closed dual one-forms, i.e. 
closed $(N-1)$-forms. In particular, in $N=1$ dimension, a closed dual 
one-form is a closed zero-form is a constant function, so the extension 
is central in this case, but not otherwise. 
However, compared to other abelian extensions, this cocycle is
very close to central, as discussed in the next section.

Modulo technicalities, $\vect(N)$ is the Lie algebra of the 
diffeomorphism group in $N$ dimensions\footnote{For this reason, I 
have denoted this algebra $diff(N)$ in previous writings.}.
Other names for this algebra are diffeomorphism
algebra and generalized Witt algebra. 
Clearly, abelian extensions of algebras of polynomial growth are more
general than central extensions of algebras of linear growth, since
central is a special case of abelian and linear is a special case of
polynomial. 

The point that the existence of
multi-dimensional Virasoro algebras is a problem for string theory's
credibility was made already in \cite{Lar97b}, in the following form:
There are no obstructions to superization, so the algebra $\vect(n|m)$ 
acting on $n|m$-dimensional superspace also admits two Virasoro-like
extensions. Na\"\i vely, one would expect to obtain such extensions of
every subalgebra of $\vect(n|m)$ by restriction, but in some boring
cases, the generically non-central extension reduces to a central one.
And in some trivial cases, it vanishes completely, apart from 
cohomologically trivial terms. The superconformal algebra is such a
boring but non-trivial case, rather than being algebraically 
distinguished.

These algebras ought to be relevant to quantum gravity, 
because\footnote{Note added in version 3: Diffeomorphism symmetry
(without compensating background fields) is the key lesson from 
general relativity. Projectivity is the key lesson from quantum
theory. To ignore the key lessons from the two main discoveries in
twentieth century physics does not seem to be the right approach to 
quantum gravity.}
\begin{itemize}
\item
The symmetry group of classical gravity is the full space-time
diffeomorphism group.
\item
In quantum theory, symmetries are only represented projectively.
On the Lie algebra level, this means that the symmetry algebra acquires
an extension (provided that the algebra is big enough).
\end{itemize}
If we put these two observations together, we see that the symmetry
algebra of quantum gravity should be an extension of the diffeomorphism
algebra $\vect(4)$. Phrased differently: I suggest that quantum gravity
should be {\em quantum general covariant}.

$\vect(N)$ has no central extensions if $N>1$
\cite{Dzhu84}, but it has many abelian extensions by irreducible modules
\cite{Dzhu96,Lar00}. However, most (possibly all) extensions by tensor 
modules are limiting cases of trivial extensions, in the sense that one
can construct a one-parameter family of trivial cocycles reducing
to the non-trivial cocycle for a critical value of the parameter (=
the conformal weight). 
In contrast, the Virasoro-like cocycles are not limits of
trivial cocycles, because the module of closed dual forms does not 
depend on any continuous parameter. 
Moreover, these are the kinds of cocycles that arise in Fock 
representations. There are indications that an interesting class of
lowest-energy modules can only be constructed in four spacetime 
dimensions \cite{Lar01a}.

2. The classification of simple infinite-dimensional Lie superalgebras 
of vector fields was recently completed by Kac, Leites, Shchepochkina, 
and Cheng. In particular, the exception $\mb(3|8)$ ($=E(3|8)$)\footnote{
When an algebra is first mentioned, I give both the names used by 
Shchepochkina and Leites (lowercase fraktur) and those used by Kac and
Cheng (uppercase roman). Subsequently, I prefer the former notation,
both because it was Shchepochkina who discovered the exceptions, and 
because Kac' notation is unsuitable to exhibit the family structure 
of the regradings.} is the deepest possible 
symmetry, at least in a technical sense (depth $3$ in its consistent 
Weisfeiler grading). Moreover, its grade zero subalgebra, which largely 
governs the representation theory, is $\ssg$, i.e. the 
non-compact form of the symmetries of the standard model. The irreps of 
an algebra of vector fields is 1-1 with the irreps of its grade zero 
subalgebra. E.g., the $\vect(N)$ irreps (tensor densities and closed 
forms) are 1-1 with the irreps of its grade zero subalgebra $gl(N)$.
One may therefore speculate that an $\mb(3|8)$ symmetry could be mistaken 
experimentally for an $\ssg$ symmetry.

It is remarkable that the sole requirements of simplicity and maximal 
depth
immediately lead to the symmetries of the standard model, without any
arbitrary symmetry breaking, compactification, magic, or mystery.
Of course, this observation does not prove that $\mb(3|8)$ is relevant
to physics, but I believe that this is line of research worth pursuing.
Although $\mb(3|8)$ acts on $3|8$-dimensional superspace, it is not
technically a supersymmetry, which would require that the fermionic 
coordinates carry Lorentz spin. 
Supersymmetry has attracted much interest because it is the only known
way to combine the Poincar\'e algebra with internal symmetries in a
non-trivial way, thus circumventing the Coleman-Mandula theorem. 
It also makes it possible to unite bosons and fermions into the 
same multiplet. On the other hand, the 
assumption that such unification is desirable remains experimentally
unproven, despite 30 years of effort. 
In contrast, in conformal field theory (CFT), several Virasoro 
multiplets (both bosons and fermions) are combined 
into a single algebraic entity united by fusion 
rules. E.g., the Ising model consists of three Virasoro modules (unity, 
spin, and energy), and the situation is similar for all other CFTs.
Avoiding supersymmetry has the
additional benefit that the theory will not be plagued by supersymmetric 
partners, none of which has been experimentally observed.

Kac has appearently been talking about the possible connection between the
exceptional superalgebras and the standard model for well over a 
year\footnote{ Kac has mainly worked with another exception
$\vle(3|6)$ ($=E(3|6)$), which has the same grade zero subalgebra but 
is not quite as deep as $\mb(3|8)$ (depth $2$ versus depth $3$)}
\cite{Kac99}, but so far has the physics community avoided to notice this 
development. Even though Kac' suggestion is not at all convincing in its 
details, the observation that the standard model algebra arises naturally 
in a mathematically deep context should evoke interest. To my knowledge,
the exceptional Lie superalgebras is the only place were this happens
in an unambigous way. In section 3 I discuss some immediate consequences of
a tentative $\mb(3|8)$ gauge theory: several generations of fermions,
absense of proton decay, no additional gauge bosons, manifest CP
violation, and particle/anti-particle asymmetry. Further 
progress must await the development of $\mb(3|8)$ representation theory.

It is an undeniable fact that some of the methods of string theory
have found very fruitful applications in other places. In particular, 
CFT is an extremely successful theory of 
two-dimensional critical phenomena \cite{BPZ84,FMS96}, and many string
theorists have made important contributions to this area. However,
that CFT is useful in two-dimensional statistical physics does
not mean that it has anything to do with four-dimensional high-energy
physics. Moreover, CFT 
is not the ToE even in statistical physics, because it does not
apply to the experimentally more interesting case of three-dimensional 
phase transitions. It should be noted that anomaly cancellation is not
an issue in this context, because every interesting statistical model 
corresponds to a CFT with non-zero central charge.

Mirror symmetry also led to progress in algebraic geometry quite a few
years back, although I know very little about this subject. However, it
should be noted that the structures described in the present paper are
local. Interesting new local structures are not so easy to come across,
and local is a prerequisite for global.

Several string theorists have informed me that symmetry considerations
not apply to M-theory, because nobody knows the algebraic structures 
behind it, and symmetries are not important in M-theory anyway. 
In all established and successful physical theories,
such as special and general relativity, Maxwell/Yang-Mills theory,
Dirac equation, and the standard model, symmetries are absolutely
fundamental. And the arguably most successful theory of all
in its domain of validity, namely CFT applied to two-dimensional 
critical phenomena, is nothing but representation theory thinly veiled
by physics formalism. 
That symmetries should be unimportant in M-theory, the
alledged mother of all theories, does not seem plausible to me. It is also
notable that the importance of symmetries was down-played during
the 1990s, at the same time as it became obvious that string 
theorists are no longer in touch with the research frontier on
infinite-dimensional Lie (super)algebras.

The multi-dimensional Virasoro algebra and its Fock modules are 
described in section 2, and the simple Lie superalgebras in section 3.
In the last section I discuss to what extent every conceivable symmetry 
is known. 
The title of this paper is thus not only a travesty of string theory's 
claim to be the ToE, but it also carries some algebraic substance.

\section{Multi-dimensional Virasoro algebra}

To make the connection to the Virasoro algebra very explicit, I write 
down the brackets in a Fourier basis. Start with the Virasoro algebra 
$Vir$:
\[
[L_m, L_n] = (n-m)L_{m+n} - {c\/12} (m^3-m) \delta_{m+n},
\]
where $\delta_m$ is the Kronecker delta. When $c=0$, 
$L_m = -i \exp(imx) d/dx$, $m \in \ZZ$. The element $c$ is central, 
meaning 
that it commutes with all of $Vir$; by Schur's lemma, it can therefore 
be considered as a c-number.
Now rewrite $Vir$ as
\bes
[L_m, L_n] &=& (n-m)L_{m+n} - c m^2 n S_{m+n}, \nl
{[}L_m, S_n] &=& (n+m)S_{m+n}, \nl
{[}S_m, S_n] &=& 0, \nl
m S_m &=& 0.
\eens
It is easy to see that the two formulations of $Vir$ are equivalent
(I have absorbed the linear cocycle into a redefinition of $L_0$).
The second formulation immediately generalizes to $N$ dimensions.
The generators are $L_\mu(m) = -i \exp(i m_\rho x^\rho) \dmu$ and
$S^\mu(m)$, where $x = (x^\mu)$, $\mu = 1, 2, ..., N$ is a point in 
$N$-dimensional space and $m = (m_\mu)$. The Einstein convention is used 
(repeated indices, one up and one down, are implicitly summed over). 
The defining relations are
\bes
[L_\mu(m), L_\nu(n)] &=& n_\mu L_\nu(m+n) - m_\nu L_\mu(m+n) \nl 
&&  + (c_1 m_\nu n_\mu + c_2 m_\mu n_\nu) m_\rho S^\rho(m+n), \nl
{[}L_\mu(m), S^\nu(n)] &=& n_\mu S^\nu(m+n)
 + \delta^\nu_\mu m_\rho S^\rho(m+n), \nl
{[}S^\mu(m), S^\nu(n)] &=& 0, \nl
m_\mu S^\mu(m) &=& 0.
\eens
This is an extension of $\vect(N)$ by the abelian ideal with basis 
$S^\mu(m)$. This algebra is even valid globally on the 
$N$-dimensional torus $\TT^N$.
Geometrically, we can think of $L_\mu(m)$ as a vector field 
and $S^\mu(m)$ as a dual one-form; the last condition expresses closedness. 
The cocycle proportional to $c_1$ was discovered by 
Rao and Moody \cite{RM94}, and the one proportional to $c_2$ by
myself \cite{Lar91}. There is also a similar multi-dimensional 
generalization of affine Kac-Moody algebras. The relevant cocycle was
presumably first written down by Kassel \cite{Kas85}, and its
modules were studied in \cite{MRY90,RMY92,Bil97,BB98,BBG01}. 
In the mathematics literature, the multi-dimensional Virasoro and 
affine algebras are often refered to as ``Toroidal Lie algebras''.

The $\vect(N)$ module spanned by $S^\nu(n)$
contains the trivial submodule $\CC^N = H^1_{dR}(\TT^N)$ (de Rham
homology) spanned by $S^\nu(0)$: 
\[
[L_\mu(m), S^\nu(0)] = \dlt^\nu_\mu m_\rho S^\rho(m) \equiv 0.
\]
Since $S^\nu(0)$ is central, it can be considered as a c-number.
If we replace $S^\nu(n) \mapsto S^\nu(n) + n_\rho F^{\nu\rho}(n)$,
where $F^{\rho\nu}(n) = -F^{\nu\rho}(n)$ and
\[
[L_\mu(m), F^{\nu\rho}(n)] = n_\mu F^{\nu\rho}(m+n)
 + \delta^\nu_\mu m_\si F^{\si\rho}(m+n)
 + \delta^\rho_\mu m_\si F^{\nu\si}(m+n),
\]
the $LS$ bracket is unchanged, whereas the cocycle becomes
\[
(c_1 m_\nu n_\mu + c_2 m_\mu n_\nu) (m_\rho S^\rho(m+n)
 + m_\rho n_\si F^{\rho\si}(m+n)).
\]
We can use this freedom to set $S^\nu(n) = 0$ for all $n\neq0$,
so we have almost a central extension
by the $N$-dimensional module $H^1_{dR}(\TT^N)$. 
However, not quite, because the condition is not preserved:
\[
[L_\mu(-n), S^\nu(n)] = -\dlt^\nu_\mu n_\rho S^\rho(0),
\]
which is non-zero unless all $S^\rho(0) = 0$. Nevertheless, this
argument shows that the multi-dimensional Virasoro cocycle is
close to central. Similarly, I expect that
$\vect(M_N)$, where $M_N$ is an $N$-dimensional manifold, has
Virasoro-like extensions labelled by $H^1_{dR}(M_N)$.

The theory of Fock modules was constructed in \cite{Lar98}.
$\vect(N)$ is generated by Lie derivatives $\Lxi$, where 
$\xi=\xmu(x)\dmu$ is a vector field; $L_\mu(m)$ is the Lie derivative 
corresponding to $\xi = -i \exp(i m_\rho x^\rho) \dmu$.
The classical modules are tensor densities \cite{Rud74} (primary fields 
in CFT parlance), which transform as
\[
[\Lxi, \phi(x)] = -\xmu(x)\dmu\phi(x)
- \dnu\xmu(x)T^\nu_\mu\phi(x),
\]
where $T^\mu_\nu$ satisfies $gl(N)$:
\[
[T^\mu_\nu, T^\rho_\si] = 
 \dlt^\rho_\nu T^\mu_\si - \dlt^\mu_\si T^\rho_\nu.
\]
Na\"\i vely, one would start from a classical field and introduce 
canonical momenta $\pi(x)$ satisfying $[\pi(x), \phi(y)] = \dlt^N(x-y)$.
This gives the following expression for $\Lxi$:
\[
\Lxi = \int d^N\,x\ \xmu(x)\pi(x)\dmu\phi(x)
+ \dnu\xmu(x)\pi(x)T^\nu_\mu\phi(x).
\]
To remove an infinite vacuum energy, we must normal order. However,
this approach only works when $N=1$, because in higher dimensions
infinities are encountered. This is in accordance with the well-known
fact the $\vect(N)$ only admits central extensions when $N=1$.

Instead, the crucial idea is to first expand all 
fields in a multi-dimensional Taylor series around the points along a 
one-dimensional curve (``the observer's trajectory''), and then to 
truncate at some finite order $p$. 
Let $\mm = (m_1, \ab m_2, \ab ..., \ab m_N)$, all $m_\mu\geq0$, be a 
multi-index of length $|\mm| = \sum_{\mu=1}^N m_\mu$. 
Denote by $\mu$ a unit vector in the $\mu$:th direction, so that
$\mm+\mu = (m_1, \ab ...,m_\mu+1, \ab ..., \ab m_N)$, and let
\[
\phim(t) = \d_\mm\phi(q(t),t)
= \underbrace{\d_1 .. \d_1}_{m_1} .. 
\underbrace{\d_N .. \d_N}_{m_N} \phi(q(t))
\]
be the $|\mm|$:th order derivative of $\phi(x)$ on the
observer's trajectory $\qmu(t)$. Such objects transform as
\bes
[\Lxi, \phim(t)] &=& \d_\mm([\Lxi,\phi(q(t))]) 
+ [\Lxi,\qmu(t)]\dmu\d_\mm\phi(q(t)) \nl
&=& -\sum_{|\nn|\leq|\mm|\leq p} T^\nn_\mm(\xi(q(t))) \phin(t), \nl
{[}\Lxi, \qmu(t)] &=& \xmu(q(t)),
\eens
where explicit expressions for the matrices $T^\nn_\mm(\xi)$ are given
in \cite{Lar98}. We thus obtain a realization of 
$\vect(N)$ on the space of trajectories in the space of tensor-valued
$p$-jets. This space consists of finitely many functions of a single
variable, which is precisely the situation where the normal ordering
prescription works. After normal ordering, we obtain a Fock
representation of the multi-dimensional Virasoro algebra described
above. The expression for $\Lxi$ reads
\[
\Lxi = \int dt\ \Big\{ \no{\xmu(q(t)) \pmu(t)} +
\sum_{|\nn|\leq|\mm|\leq p}
\no{ \pim(t) T^\nn_\mm(\xi(q(t))) \phin(t)  } \Big\},
\]
where $[\pnu(s), \qmu(t)] = \dlt^\mu_\nu \dlt(s-t)$ and
$[\pim(s), \phin(t)] = \dlt^\mm_\nn \dlt(s-t)$.

Some observations are in order:
\begin{enumerate}
\item
The action on jet space is non-linear; the observer's trajectory 
transforms non-linearly, and although vector fields act linearly on
the Taylor coefficients, they act with matrices depending non-linearly 
on the base point. Hence the resulting extension is non-central.
\item
Classically, $\vect(N)$ acts in a highly reducible fashion. In fact, the 
realization is an infinite direct sum because neighboring points on the 
trajectory transform independently of each other. To lift this 
degeneracy, I introduced an additional $\vect(1)$ factor, describing 
reparametrizations. The relevant algebra is thus the DRO (Diffeomorphism, 
Repara\-metri\-zation, Observer) algebra $DRO(N)$, which is the extension
of $\vect(N)\oplus\vect(1)$ by its four Virasoro-like cocycles.
\item
The reparametrization symmetry can be eliminated with a constraint, but
then one of the spacetime direction (``time'') is singled out. Two of
the four Virasoro-like cocycles of $DRO(N)$ transmute into the complicated
anisotropic cocycles found in \cite{Lar97a}; these are colloquially known
as the ``messy cocycles''. By further specialization to scalar-valued
zero-jets on the torus, the results of Rao and Moody are recovered 
\cite{RM94}.
\end{enumerate}

After the Fock modules were constructed, more interesting lowest-energy
modules were considered in \cite{Lar99}. This unpublished paper is far
too long and contains some flaws, mainly because I didn't have the right
expressions for the abelian charges\footnote{ I refer to the parameters
multiplying the cocycles as {\em abelian charges}, in analogy with the 
central charge of the Virasoro algebra.} at the time, but I think that 
the main idea is sound. 

In classical physics one wants to find the stationary surface $\Sigma$,
i.e. the set of solutions to the Euler-Lagrange (EL) equations, 
viewed as a submanifold embedded in configuration space $\QQ$. Dually,
one wants to construct the function algebra $C(\Sigma) = C(\QQ)/\II$,
where $\II$ is the ideal generated by the EL equations. For each field
$\fa$ and EL equation $\Ea=0$, introduce an anti-field $\fsa$ of 
opposite Grassmann parity. The extended configuration space 
$C(\QQ^*) = C[\phi,\fs]$ can be decomposed into subspaces $C^g(\QQ^*)$ 
of fixed antifield number $g$, where $\afn\fa = 0$, $\afn\fsa = 1$.
The Koszul-Tate (KT) complex 
\[
0 \larroww \dlt C^0(\QQ^*) \larroww \dlt C^1(\QQ^*) \larroww \dlt 
C^2(\QQ^*) \larroww \dlt \ldots,
\]
where $\dlt\fa = 0$ and $\dlt\fsa = \Ea$, yields a resolution of 
$C(\Sigma)$; the cohomology groups $H^g(\dlt) = 0$ unless $g=0$, and 
$H^0(\dlt) =C(\QQ)/\II$ \cite{HT92}.

My idea was to consider not just functions on the stationary surface, 
but all differential operators on it. The KT differential $\dlt$ can
then be written as a bracket: $\dlt F = [Q,F]$, where the KT charge 
$Q = \int \Ea\psa$ and $\psa$ is the canonical momentum corresponding
to $\fsa$. If we pass to the space of $p$-jets before momenta are
introduced, the construction of Fock modules above applies. Since the
KT charge consists of commuting operators, it does not need to be 
normal ordered, and the cohomology groups are well-defined $DRO(N)$
modules of lowest-energy type.

I think that this construction can be viewed as a novel method for 
quantization, although the relation to other methods is not clear. 
However, it was never my intention to invent a new quantization scheme,
but rather to construct interesting $DRO(N)$ modules. An outstanding
problem is to take the maximal jet order $p$ to infinity, because
infinite jets essentially contain the same information as the original
fields. This limit is problematic, because the abelian charges diverge
with $p$, but it seems that this difficulty may be bypassed in four
dimensions, as announced in \cite{Lar01a}.

\section{ 
Classification of simple infinite-dimensional Lie superalgebras of 
vector fields and the exceptional Lie superalgebra $\mb(3|8)$ }

Technically, the classification deals with polynomial vector fields 
acting on a superspace of dimension $n|m$ ($n$ bosonic and $m$ fermionic 
directions), i.e. of simple subalgebras of $\vect(n|m)$. However, the 
restriction to polynomials is not philosophically essential, because 
the results also apply to functions that can be approximated by 
polynomials, e.g. analytic functions. So it is really a classification 
of simple algebras of local vector fields. 

Let $\oj \subset vect(n|m)$ be such an algebra. It has a Weisfeiler 
grading of depth $d$ if it can be written as
\[
\oj = \oj_{-d} + ... +\oj_{-1} + \oj_0 + \oj_1 + ...,
\]
where $\oj_{-1}$ is an irreducible $\oj_0$ module and
$\oj_k$ consists of vector fields that are homogeneous of degree $k$.
However, it is not the usual kind of homogeneity, because we do not 
assume that all directions are equivalent. Denote the coordinates of 
$n|m$-dimensional superspace by $x^i$ and let $\d_i$ be the 
corresponding derivatives. Then we define the grading by introducing 
positive integers $w_i$ such that $\deg x^i = w_i$ and $\deg \d_i = -w_i$.
The operator which computes the Weisfeiler grading is 
$Z = \sum_i w_i x^i \d_i$, and $\oj_k$ is the subspace of vector fields 
$X = X^i(x) \d_i$ satisfying $[Z, X] = k X$. If we only considered $\oj$
as a graded vector space, we could of course make any choice of integers 
$w_i$, but we also want $\oj$ to be graded as a Lie algebra:
$[\oj_i, \oj_j] \subset \oj_{i+j}$. 
The depth $d$ is identified with the maximal $w_i$.
Denote the negative part by $\oj_- =  \oj_{-d} + ... +\oj_{-1}$; it is
a nilpotent algebra and a $\oj_0$ module.

The main tool for constructing algebras of vector fields is Cartan 
prolongation. In the mathematics literature, it is defined recursively in
a way which is not so easy for a physicist to understand.
Therefore, I propose the following alternative definition of Cartan 
prolongation:
\begin{enumerate}
\item
Start with a realization for the non-positive part $\oj_0\ltimes\oj_-$
of $\oj$ in $n|m$-dimensional superspace.
\item
Determine the most general set of structures preserved by 
$\oj_0\ltimes\oj_-$. Such a structure is either some differential form,
or an equation satisfied by forms (Pfaff equation), or a system of Pfaff
equations.
\item
Define the Cartan prolong $\oj = (\oj_{-d}, ..., \oj_{-1}, \oj_0)_*$
as the full subalgebra of $\vect(n|m)$ preserving the same structures.
\end{enumerate}
Clearly, the set of vector fields that preserve some structure 
automatically define a subalgebra of $\vect(n|m)$. By choosing the
maximal set of equations, this subalgebra must be $\oj$ itself.
If one is lucky, $\oj$ is now simple and infinite-dimensional. The 
mathematicians have determined when this happens. 

The simplest example is the prolong $(\brep n, sl(n))_*$, 
where $\brep n$ stands for the $n$-dimensional $sl(n)$ module with 
basis $\d_i$. The vector fields are, at non-positive degrees
\renewcommand{\arraystretch}{1.4}
\[
\begin{array}{|r|l|}
\hline
\deg&\hbox{vector field} \\
\hline
-1:&\d_i \\
0:  &x^i \d_j - {1\/n} \delta^i_j x^k \d_k \\
\hline
\end{array}
\]
\renewcommand{\arraystretch}{1.1}
These vector fields $X = X^i(x)\d_i$ preserve the volume form
$vol$. The prolong is the algebra $\svect(n)$ of
all divergence-free vector fields, all of which preserve $vol$.

The classification \cite{Kac98,Kac99} consists of a list of ten 
series\footnote{ Leites and Shchepochkina use different names for
$\sko_\beta(n)$, $\widetilde\sle(n)$ and $\widetilde\sko(n)$, but
unfortunately I do not quite understand their notation.}:
\[
\begin{array}{ll}
\vect(n|m) = W(n|m)
&\hbox{ arbitrary v.f. in $n|m$ dimensions, }\\
\svect(n|m) = S(n|m)
&\hbox{ divergence-free v.f., }\\
{\mathfrak h}(n|m) = H(n|m)
&\hbox{ Hamiltonian v.f. ($n$ even), }\\
\fle(n) = HO(n|n)               
&\hbox{ odd Hamiltonian or Leitesian v.f. $\subset \vect(n|n)$, }\\
\sle(n) = SHO(n|n)
&\hbox{ divergence free Leitesian v.f., }\\
\kk(n|m) = K(n|m)
&\hbox{ contact v.f. ($n$ odd), }\\
\ko(n) = KO(n|n+1)
&\hbox{ odd contact v.f. $\subset \vect(n|n+1)$, }\\
\sko_\beta(n) = SKO(n|n+1;\beta) 
&\hbox{ a deformation of div-free odd contact v.f.,} \\
\widetilde\sle(n) = SHO\til(n|n)
&\hbox{ a deformation of $\sle(n)$}, \\
\widetilde\sko(n) = SKO\til(n|n+1)
&\hbox{ a deformation of $\sko(n)$}.
\\
\end{array}
\]
Moreover, there are five exceptions, described as Cartan prolongs
\[
\begin{array}{l}
\vas(4|4) = E(4|4)
 = ({\rm spin}, {\mathfrak {as}})_*, \\
\vle(3|6) = E(3|6) 
 = (\brep 3, \brep3^*\!\otimes\!\brep2, \ssg)_*, \\
\ksle(5|10) = E(5|10) 
 = (\brep 5, \brep5^*\!\wedge\!\brep5^*, sl(5))_*, \\
\mb(3|8) = E(3|8) 
 = (\brep 2, \brep 3, \brep 3^*\!\otimes\!\brep 2, \ssg)_*,\\ 
\kas(1|6) = E(1|6)
 \subset \kk(1|6) = (\brep 1, \brep 6, so(6)\!\oplus\!g1(1)). 
\end{array}
\]
The finite-dimensional superalgebra ${\mathfrak {as}}$ is the central
extension of the special periplectic algebra ${\mathfrak {spe}}(4)$ 
discovered by A. Sergeev; ${\mathfrak {spe}}(n)$ has no central 
extensions for other values of $n$.
The construction of $\kas(1|6)$ is slightly more complicated than Cartan
prolongation; it is a subalgebra of $\kk(1|6)$. 

A more geometric way to describe these algebras is by stating what
structures they preserve, or what other conditions the vector fields obey. 
This is a very new result for the exceptions \cite{Lar01b}.
\renewcommand{\arraystretch}{1.4}
\[
\begin{array}{|l|l|l|}
\hline
\hbox{Algebra} & \hbox{Basis} & \hbox{Description/structure preserved}\\
\hline
\vect(n|m) & u^i, \th^a & -\\
\svect(n|m) & u^i, \th^a & vol\\
{\mathfrak h}(n|m) & u^i, \th^a
& \om_{ij}du^idu^j + g_{ab}d\th^ad\th^b \\
\fle(n) & u^i, \th_i & du^id\th_i \\
\sle(n) & u^i, \th_i & du^id\th_i, vol \\
\kk(n+1|m) & t, u^i, \th^a
& dt + \om_{ij}u^idu^j + g_{ab}\th^ad\th^b = 0 \\
\ko(n) & \tau, u^i, \th_i
& d\tau + u^id\th_i + \th_idu^i = 0 \\
\sko_\beta(n)& \tau, u^i, \th_i
& M_f \in \ko(n): \div_\beta M_f = 0 \\
\widetilde\sle(n) & u^i, \th_i
& (1+\th_1..\th_n)\xi, \xi \in \sle(n)\\
\widetilde\sko(n) & \tau, u^i, \th_i
& (1+\th_1..\th_n)\xi, \xi \in \sko_{n+2\/n}(n)\\
\ksle(5|10) & u^i, \th_{ij}
&du^i + {1\/4}\eps^{ijklm}\th_{jk}d\th_{lm} = 0, vol\\
\vle(3|6) & u^i, \th_{ia}
&du^i + \eps^{ijk}\eps^{ab}\th_{ja}d\th_{kb} =0, vol \\
\kas(1|6) & t, \th^a 
& K_f \in \kk(1|6): {\d^3 f\/\d\th^c\d\th^c\d\th^c}
= \eps_{abcdef}{\d^3 f\/\d\th_d\d\th_e\d\th_f} (?) \\
\hline
\end{array}
\]
In this table, $u^i$ denotes bosonic variables, $\th^a$, $\th_i$
and $\th_{ij}$ fermionic variables, and $t$ ($\tau$) is an extra bosonic
(fermionic)
variable. The indices range over the dimensions indicated: $i=1,...,n$
and $a=1,...,m$, except for $\vle(3|6)$ where $a=1,2$ only. 
$\th_{ij}=-\th_{ji}$ in $\ksle(5|10)$ so there are only ten independent
fermions. $\om_{ij} = -\om_{ji}$ and $g_{ab} = g_{ba}$ are structure
constants, and $\eps^{ab}$, $\eps^{ijk}$ and $\eps^{ijklm}$ are the totally
anti-symmetric constant tensors in the appropriate dimensions.
The notation $\al=0$ (or $\al^i=0$) implies that it is this Pfaff 
equation that is preserved, not the form $\al$ itself; 
a vector field $\xi$ acts on $\al$ as $\L_\xi\al = f_\xi\al$, $f_\xi$ some
polynomial function.
$vol$ denotes the volume form; vector fields preserving $vol$ satisfy
$\div\, \xi = (-)^{\xi\mu + \mu}\dmu\xmu = 0$. 
\[
\div_\beta M_f = 2(-)^f ({\d^2f\/\d u^i\d\th_i} +
\big( u^i{\d\/\d u^i} + \th_i{\d\/\d \th_i} - n\beta){\d f\/\d\tau} \big)
\]
is a deformed divergence.
In the conjectured description of $\kas(1|6)$, indices are lowered by
means of the metric $g_{ab}$. The geometrical meaning of $\vas(4|4)$ is 
not clear to me, but the differential equations that it satisfies are
written down in \cite{Sh99}.

Finally, let us describe the exceptional algebra
$\mb(3|8) = (\brep 2, \brep 3, \brep3^*\!\otimes\!\brep 2, \ssg)_*$.
A basis for $3|8$-dimensional space is given by 
\[
\begin{array}{ll}
\th_{ia}, &\hbox{(degree $1$, $2\times3 = 6$ fermions)},\\
u^i,  &\hbox{(degree $2$, $3$ bosons)},\\
\vth^a,  &\hbox{(degree $3$, $2$ fermions)},
\end{array}
\]
where $i=1,2,3$ is a three-dimensional index and $a=1,2$ a two-dimensional 
index. 
Denote the corresponding derivatives by $d^{ia}$, $\d_i$ and 
$\eth_a$, respectively. 
We can explicitly describe the vector fields at non-positive 
degree\footnote{Note added in version 3: The following two formulas 
were incorrect in version 2.}.
\[
\begin{array}{|r|ll|}
\hline
\deg&\hbox{vector field} &\\
\hline
-3 & F_a = \eth_a &\\
-2 & E_i = \d_i + \th^a_i\eth_a &\\
-1 & D^{ia} = d^{ia} + 3\eps^{ijk}\th^a_j\d_k
+ \eps^{ijk}\th^a_j\th^b_k\eth_b + u^i\eth^a &\\
0& I^k_l = u^k \d_l - \th_{la} d^{ka} 
 - {1\/3}\dlt^k_l( \ud - \thd ) & (sl(3))\\
0& J^c_d = \vth^c \eth_d - \th_{id} d^{ic} 
- {1\/2}\dlt^c_d( \veth - \thd ) & (sl(2))\\
0& Z = 3 \veth + 2 \ud + \thd & (gl(1)) \\
\hline
\end{array}
\]
Geometrically, $\mb(3|8)$ preserves the system of dual Pfaff equations 
$\tilde D^{ia} = 0$, where
\bes
\tilde D^{ia} &=& d^{ia} - 3\eps^{ijk}\th^a_j\d_k
+ \eps^{ijk}\th^a_j\th^b_k\eth_b - u^i\eth^a.
\eens
In other words, $\mb(3|8)$ consists of vector fields $X$ such that
$[\tilde D^{ia}, X] = f^{ia}_{jb}(X)\tilde D^{jb}$ for some polynomial
functions $f^{ia}_{jb}(X)$.

Given an algebra $\oj$, a subalgebra $\hh \subset \oj$, and an $\hh$
representation $\rep$, one can always construct the induced $\oj$
representation $\U(\oj) \otimes_{\U(h)} \rep$, where $\U(\cdot)$ denotes 
the universal enveloping algebra. For finite-dimensional algebras, the 
induced representation is usually too big to be of interest, but for 
Cartan prolongs the situation is different. There is a 1-1 
correspondance between $\oj$
irreps and irreps of its grade zero subalgebra $\oj_0$, as follows.
Start from a $\oj_0$ irrep and construct the corresponding induced 
$\oj$ representation. If this is irreducible, which is often the case, 
we are done. Otherwise, it contains an irreducible subrepresentation.
A well-known example is the Cartan prolong 
$\vect(n) = (\brep{n}, gl(n))_*$. A $gl(n)$ module is a tensor with 
certain symmetries, and the induced $\vect(n)$ module is the 
corresponding tensor density. It is irreducible unless it is totally 
anti-symmetric and has weight zero, i.e. it is a differential form. 
Then it contains the submodule of closed forms \cite{Rud74}.

It sometimes happens that a Lie superalgebra can be realized in more
than one way as vector fields acting on superspaces (of different 
dimensions). One then says that the algebra has a Weisfeiler regrading. 
No proper Lie algebra has a regrading. The five exceptional Lie 
superalgebras have fifteen regradings altogether, listed in the format
$name(n|m):depth$: 
$\vle(4|3):1$, $\vle(5|4):2$, $\vle(3|6):2$, $\vas(4|4):1$,
$\kas(1|6):2$, $\kas(5|5):2$, $\kas(4|4):1$, $\kas(4|3):1$,
$\mb(4|5):2$, $\mb(5|6):2$, $\mb(3|8):3$,
$\ksle(9|6):2$, $\ksle(11|9):2$, $\ksle(5|10):2$, $\ksle(11|9;CK):3$.
Note that $\ksle(5|10)$ has two inequivalent realizations on 
$11|9$-dimensional superspace.

Among all gradings, the consistent ones play a distinguished role. A
grading is consistent if the even subspaces are purely bosonic and the odd
ones are purely fermionic. Most gradings are inconsistent, and no 
superalgebra has more than one consistent grading.

\smallskip\noindent
{\bf Theorem \cite{Kac98}}: The only simple Lie
superalgebras with consistent gradings are the contact algebras
$\kk(1|m)$ and the exceptions $\ksle(5|10)$, $\vle(3|6)$, 
$\mb(3|8)$ and $\kas(1|6)$. 
\smallskip

If we change the function class from polynomials to Laurent polynomials, 
the contact algebras $\kk(1|m)$ for $m\leq 4$ have central extensions, 
known in physics as the $N=m$ superconformal algebra (the case $m=0$ is 
the Virasoro algebra). 
Conjecture: these central extensions are obtained by restriction from the 
abelian Virasoro-like extensions of $\vect(1|m)$. This statement has
been proven in the case $m=1$ \cite{Lar97b}. 
$\kk(1|m)$ does not have central extensions for $m > 4$,
but clearly it has abelian Viraosoro-like extensions, which could be called
the $N=m$ superconformal algebra also for $m>4$.
Although the superconformal algebras play an important role in
string theory, their description as the vector fields that preserve the
Pfaff equation $\al = dt + g_{ab}\th^ad\th^b = 0$
might be unknown to some physicists. Nor is it common knowledge that the
restriction to $m\leq4$ is unnecessary in the centerless case.

If $\mb(3|8)$ is to replace the standard model algebra, it must be gauged,
i.e. one must pass to the algebra $\map(4,\mb(3|8))$ of maps from 
four-dimensional spacetime to $\mb(3|8)$. After inclusion of gravity,
the full symmetry algebra becomes $\vect(4)\ltimes\map(4,\mb(3|8))$.
Obviously,
\[
\vect(4)\ltimes\map(4,\mb(3|8)) \subset
\vect(4)\ltimes\map(4,\vect(3|8)) \subset
\vect(7|8).
\]
The middle algebra consists of those vector fields in $7|8$-dimensional 
space that preserve the splitting between horizontal and vertical
directions. This chain of inclusions proves that 
$\vect(4)\ltimes\map(4,\mb(3|8))$ has well-defined abelian extensions and
Fock modules; consider the restriction from $\vect(7|8)$.

The algebra $\map(4,\oj)$ encodes a very natural generalization of the
gauge principle. Gauging a rigid symmetry $\oj_0$ makes it local in
spacetime, but it is still rigid in the fiber directions; the 
finite-dimensional algebra $\oj_0$ acts on each fiber. The replacement
of $\oj_0$ by its infinite-dimensional prolong $\oj = (\oj_-, \oj_0)_*$
makes the symmetry local in the fibers as well, while maintaining the
essential features of representations and physical predictions. The only
freedom lies in the nilpotent algebra $\oj_-$; it is natural to choose it
such that $\oj$ is simple.

Unfortunately,
the details of an $\mb(3|8)$ gauge theory must await the development of 
the $\mb(3|8)$ representation theory, in particular the list of  
degenerate irreducible modules which is not yet available.\footnote{Note
added in version 3: This list has now been worked out by several authors;
the first were V.G. Kac and A.N. Rudakov:
  {\it Complexes of modules over exceptional Lie superalgebras $E(3|8)$ 
  and $E(5|10)$}, {\tt math-ph/0112022} (2001). 
As explained in \cite{Lar01c}, the hypercharge assignments did not quite
work out the way I hoped.}
However, the analogous list 
for $\vle(3|6)$ has recently been worked out by Kac and Rudakov 
\cite{KR00a,KR00b}, so a $\vle(3|6)$ gauge theory can be written down at 
this time \cite{Lar01c}. But even before its construction, we know enough
about the exceptional Lie superalgebras to make some non-trivial, and 
falsifyable, predictions.

Let me sketch the construction of a $\oj$ gauge theory, where 
$\oj=\vle(3|6)$ or $\oj=\mb(3|8)$. Let $\rep$ denote a $\oj_0=\ssg$ 
module. The $\oj$ tensor modules $T(\rep)$ are
$\rep$-valued functions $\psi(\th,u,\vth)$, where 
$(\th,u,\vth) \in \CC^{3|8}$ are the coordinates in the internal 
directions ($(\th,u) \in \CC^{3|6}$ for $\vle(3|6)$).
The corresponding $\map(4,\oj)$ tensor module, also denoted by 
$T(\rep)$, is either irreducible, or contains an 
irreducible submodule consisting of tensors satisfying
$\nabla\psi(x,\th,u,\vth)=0$. Here 
$x = (x^\mu) \in \RR^4$ is a point in spacetime and 
$\nabla$ is a morphism inherited from $\oj$, i.e. a differential operator
acting on the internal directions only. Kac and Rudakov call the modules 
$I(\rep) = \ker\nabla = \im\nabla'$ (cohomology is almost always absent)
``degenerate irreducible modules'', but I will use the shorter name
{\em form modules}\footnote{More precisely, one should call $T(\rep)$ a 
form module if it is reducible, and its irreducible quotient 
$I(\rep) = \ker\nabla \subset T(\rep)$ a {\em closed}
form module. The distinction should be clear from the context.}, 
because we can think of $\psi$ as a differential 
form and of $\nabla$ as the exterior derivative.

\smallskip\noindent
{\bf Theorem \cite{KR00a,KR00b}}: The $\vle(3|6)$ form modules are
\[
\begin{array}{l}
\Omega_A(p,r) = I(p,0;r;{2\/3} p-r), \\
\Omega_B(p,r) = I(p,0;r;{2\/3} p+r+2), \\
\Omega_C(q,r) = I(0,q;r;-{2\/3} q-r-2), \\
\Omega_D(q,r) = I(0,q;r;-{2\/3} q+r),
\end{array}
\]
where $(p,q;r;y)$ is an $\ssg$ lowest weight ($(p,q)\in\NN^2$ for $sl(3)$,
$r\in\NN$ for $sl(2)$, and $y\in\CC$ for $gl(1)$).
\smallskip

The gl(1) generator is the operator which computes the
Weisfeiler grading; up to normalization, it can be identified 
with weak hypercharge: $Y = Z/3$ \cite{Kac99}. By the Gell-Mann-Nishiyima
formula, the electric charge in $(p,q;r;y)$ ranges from $y/2-r/2$ to
$y/2+r/2$ in integer steps.

We identify the fermions (quarks and leptons) in the first generation with
form modules, and additionally assume that they transform as spinors
under the Lorentz group in the usual way. This leads to the following
assignment of fermions:
\renewcommand{\arraystretch}{1.2}
\[
\begin{array}{|c|c|ccc|c|} 
\hline
\hbox{Multiplet} & \hbox{Charges} &&&& \hbox{Form} \\
\hline
(0,1;1;{1\/3}) & {2\/3},-{1\/3} & 
\begin{pmatrix} u_L \\ d_L \end{pmatrix} &
\begin{pmatrix} c_L \\ s_L \end{pmatrix} &
\begin{pmatrix} t_L \\ b_L \end{pmatrix} &
\Omega_D(1,1)
\\
(1,0;1;-{1\/3}) & -{2\/3},{1\/3} & 
\begin{pmatrix} \tilde u_R \\ \tilde d_R \end{pmatrix} &
\begin{pmatrix} \tilde c_R \\ \tilde s_R \end{pmatrix} &
\begin{pmatrix} \tilde t_R \\ \tilde b_R \end{pmatrix} &
\Omega_A(1,1)
\\
(1,0;0;-{4\/3}) & -{2\/3} & \tilde u_L & \tilde c_L & \tilde t_L &-\\
(0,1;0;{4\/3}) & {2\/3} &       u_R & c_R & t_R & -\\
(0,1;0;-{2\/3}) & -{1\/3} & d_R & s_R & b_R &
\Omega_D(1,0)
\\
(1,0;0;{2\/3}) & {1\/3} & \tilde d_L & \tilde s_L & \tilde b_L & 
\Omega_A(1,0)
\\
\hline
(0,0;1;-1) & 0,-1 & 
\begin{pmatrix} \nu_{eL} \\ e_L \end{pmatrix} &
\begin{pmatrix} \nu_{\mu L} \\ \mu_L \end{pmatrix} &
\begin{pmatrix} \nu_{\tau L} \\ \tau_L \end{pmatrix} &
\Omega_A(0,1)
\\
(0,0;1;1) & 0,1 & 
\begin{pmatrix} \tilde \nu_{eR} \\ \tilde e_R \end{pmatrix} &
\begin{pmatrix} \tilde \nu_{\mu R} \\ \tilde \mu_R \end{pmatrix} &
\begin{pmatrix} \tilde \nu_{\tau R} \\ \tilde \tau_R \end{pmatrix} &
\Omega_D(0,1)
\\
(0,0;0;2) & 1 & \tilde e_L & \tilde \mu_L & \tilde \tau_L &
\Omega_C(0,0)
\\
(0,0;0;-2) & -1 & e_R & \mu_R & \tau_R &
\Omega_B(0,0)
\\
\hline
\end{array}
\]

In the usual $\oj_0 = \ssg$ gauge they, the gauge bosons are 
$\map(4,\oj_0)$ connections, which can be regarded as functions 
$A^a_\mu(x)$, with $a$ a $\oj_0$ index. {F}rom this one can construct
the covariant derivative $\dmu + A^a_\mu(x)T_a$, where $T_a$ are the
$\oj_0$ generators. In a $\oj$ gauge theory, the gauge bosons must 
analogously be taken as $\map(4,\oj)$ connections, i.e. $\oj_0$-valued
functions $A^a_\mu(x,\th,u,\vth)$ with a twisted action of $\oj$.
The $\map(4,\oj)$ covariant derivative 
$D_\mu = \dmu + A^a_\mu(x,\th,u,\vth)T_a$ acts on tensor and form
modules, making $D_\mu\psi$ transform as a vector-spinor under the Lorentz
group, and as $\psi$ itself under $\oj$.
The $\map(4,\oj)$ action on $A^a_\mu$ follows from the definition
$A^a_\mu T_a = D_\mu-\dmu$, and the curvature is 
$F_{\mu\nu} = [D_\mu,D_\nu]$, as usual.
The equations of motion are thus the same as in the standard model,
except for the replacements $\psi(x)\to\psi(x,\th,u,\vth)$, 
$A^a_\mu(x)\to A^a_\mu(x,\th,u,\vth)$, and the additional conditions
$\nabla\psi(x,\th,u,\vth)=0$.

Since fermions in several different form modules are introduced, the
natural question is which multiplets to consider. I suggest that one
must choose a fundamental set of form modules (actually two sets,
one for each helicity), from which all 
forms can be built by taking appropriate ``wedge products'',
i.e. bilinear maps 
\[
\wedge: I(\rep_1)\times I(\rep_2) \longrightarrow
I(\rep_3) \subset T(\rep_1)\otimes T(\rep_2) = T(\rep_1\otimes \rep_2),
\]
where  $\rep_3 \subset \rep_1\otimes \rep_2$ is such that $T(\rep_3)$
is reducible.
Unfortunately, these wedge products are not known, even 
for $\vle(3|6)$, so this idea remains a hypothesis.

The $\vle(3|6)$ gauge theory leads to several predictions,
which are in rough agreement with experiments:
\begin{enumerate}
\item
Since $\oj_0=\ssg$, many good properties are inherited from the standard
model.
\item
Under the restriction to $\oj_0$, a $\oj$ module $I(\rep)$ decomposes
into many $\oj_0$ modules, several of which may be isomorphic to $\rep$.
These may be identified with different generations of the same fermion.
Thus I suggest unification of e.g. $(d_R, s_R, b_R)$ into the same 
$\vle(3|6)$ multiplet $\Omega_D(1,0)$. 
However, since the isomorphic $\oj_0$ modules sit in different
ways in the parent $\oj$ module, the properties of the different 
generations should be different.
\item
The fermions in the first generation belong to different $\oj$ multiplets.
In particular, the proton is stable because quarks and leptons are not
unified into the same multiplet.
\item
The $\oj$ connections are ``fatter'' than the $\oj_0$ connections, being
functions of the internal coordinates as well, but they have no new 
components. Thus there are no new unobserved gauge bosons. This rules out
e.g. technicolor scenarios.
\item
A particle/anti-particle asymmetry has been built into the theory:
$\oj$ consists of subspaces
$\oj_k$, where the $Z$ eigenvalue $k$ ranges from $-3$ ($-2$ for 
$\vle(3|6)$) to $+\infty$, so this algebra is 
not symmetric under the reflection $Z \to -Z$. A related observation is 
that in the definition
$\mb(3|8) = (\brep 2, \brep 3, \brep 3^*\!\otimes\!\brep 2, \ssg)_*$,
the fundamental $sl(3)$ modules $\brep 3$ and $\brep 3^*$ enter 
asymmetrically. This asymmetry should be reflected in nature, maybe 
in the relative abundance of matter and anti-matter.
\item
CP amounts to the permutations 
$\Omega_A(p,r) \leftrightarrow \Omega_D(p,r)$,
$\Omega_B(p,r) \leftrightarrow \Omega_C(p,r)$,
but the directions of the morphisms $\nabla$ are unchanged.
Equivalently, if a particle is described as $\ker \nabla = \im \nabla'$,
its CP conjugate is $\im \nabla = \ker \nabla'$, i.e.
the kernel and image are interchanged.
Thus the theory predicts manifest CP violation, without the need for 
$\theta$ vacua.
However, CPT is conserved because T reverses the direction of all 
morphisms.
\item
$\vle(3|6)$ and $\mb(3|8)$ shed no light on the origin of masses.
A Higgs particle (a boson of type $I(0,0;1;1) = \Omega_D(0,1)$, i.e. an
$sl(2)$ doublet with charges $0$ and $+1$) can be added by hand in
the same way as in the standard model, but this is no more (and no less)
satisfactory than in the standard model.
\item
One could add a fermion in $\Omega_A(0,0) = \Omega_D(0,0) = I(0,0;0;0)$.
It would appear as a sterile neutrino.
\end{enumerate}

Despite these successes, the $\vle(3|6)$ theory is fundamentally flawed.
In the table above, we see that the right-handed $u$ quark
$u_R = (0,1;0;{4\/3})$ and its anti-particle 
$\tilde u_L = (1,0;0;-{4\/3})$ do not correspond to form modules.
The absence of a right-handed $u$ quark is of course fatal for the 
$\vle(3|6)$ theory. However, this result has no bearing on $\mb(3|8)$,
because its list of form modules has not yet been worked out,
and one may hope that all quarks and leptons in the table above
will correspond to forms. Conversely, we can turn this 
physical requirement into a conjecture about the list of form
modules for $\mb(3|8)$. Fortunately, this problem is of considerable
mathematical interest in its own right, so algebraists are likely to
attack it soon.

Finally some history. In 1977, Kac classified the finite-dimensional Lie 
superalgebras \cite{Kac77}, and conjectured that the infinite-dimensional 
case would be the obvious super analogoue to Cartan's list $(W_n, S_n, 
H_n, K_n)$ \cite{Car09}. However, it was immediately pointed out 
\cite{ALSh80,Lei77} that there are also odd versions of 
$H_n$ and $K_n$; the odd Hamiltonian algebra was introduced
in physics soon thereafter by Batalin and Vilkovisky. The deformations 
were also found \cite{Lei85}. It was a complete surprise when 
Shchepochkina found three exceptions \cite{Sh83}, followed by two more 
\cite{Sh87,Sh97,Sh99}; the exception $\kas(1|6)$ was independently 
found by Cheng and Kac \cite{CK97}.
Important techniques were developed by Leites and Shchepochkina 
\cite{LSh88}, and the classification was known to 
them in 1996 (announced in \cite{Sh97}), 
but I have only seen their paper in preprint form \cite{LSh00}. 
Meanwhile, Kac also worked out the classification \cite{Kac98}, together 
with Cheng \cite{CK99}.

\section{Discussion}
In the introduction, I made the claim that $\mb(3|8)$ and the 
multi-dimensional Virasoro algebra are essentially the deepest and most
general symmetries possible. This statement assumes that one interprets 
the word ``symmetry'' in a conservative sense, to mean semisimple Lie 
algebras or superalgebras, finite-dimensional algebras or 
infinite-dimensional algebras of vector fields,
and abelian extensions thereof.  One can view this list as the definition 
of the present paradigm. 
This list covers many algebraic structures, including all infinitesimal 
quantum deformations, which are always of the form
\bes
[J^a, J^b] &=& f^{ab}{}_c J^c + k^{ab}{}_i \hbar E^i + O(\hbar^2), \nl
{[}J^a, \hbar E^i] &=& g^{ai}{}_j \hbar E^j + O(\hbar^2), \nl
{[}\hbar E^i, \hbar E^j] &=& 0  + O(\hbar^2).
\eens
We see that every deformation reduces to an abelian (but not necessarily
central) extension in the $\hbar\to0$ limit. 

However, my definition of symmetry excludes algebras of exponential 
growth, e.g. non-affine Kac-Moody algebras. The reason is that I think
of algebras as infinitesimal transformation groups. In particular,
$\vect(n|m)$ is the Lie algebra of the diffeomorphism group 
in $n|m$ dimensions, and this algebra is of polynomial growth. It
is difficult to visualize any interesting symmetry that is 
essentially bigger than the full diffeomorphism group, maybe acting
on the total space of some bundle over the base manifold. 
That the lack of an explicit realization 
is a problem for non-affine Kac-Moody algebras has been drastically
formulated by Kac: ``It is a well kept secret that the
theory of Kac-Moody algebras has been a disaster.'' \cite{Kac97}.

If one stays within the present paradigm, there are really no unknown
possibilities; classifications exist\footnote{ Abelian extensions of 
superalgebras are not yet classified.}. However, if there is a 
paradigm shift, all bets are off. Paradigm shifts have occurred in the 
past. Examples are given by the transitions from Lie algebras to 
superalgebras, from semisimple algebras to central extensions, and
presumably also from central extensions to abelian but
non-central Virasoro-like extensions. 

What makes me believe that there
will be no further paradigm shift is the convergence between the
deepest algebraic structures and the deepest experimental physics: 
both mathematics and physics seem to require four dimensions and 
the $\ssg$ symmetry at their deepest levels. 
But even if these structures turn out to be unrelated to physics, it is 
still worthwhile to pursue them. Hardly anything in mathematics is more 
natural than the representation theory of interesting algebras, so the
efforts will not be wasted.

\end{document}